\documentclass[12pt,a4paper]{article}
\usepackage[utf8]{inputenc}
\usepackage{amsmath,amsopn}
\usepackage{amsfonts}
\usepackage{amssymb}
\usepackage{amsthm}
\usepackage{float}
\usepackage{url}
\usepackage{subfigure}
\usepackage{caption}
\usepackage{hyperref}
\usepackage{graphicx}
\usepackage{cite}
\usepackage{xcolor}
\hypersetup{
    colorlinks = true,
    linkcolor={green!40!black},
    citecolor=[rgb]{0.31,0.47,0.26},
    urlcolor=[rgb]{0.4,0.6,0.8}
}
\usepackage[left=2.5cm,right=2.5cm,top=2.5cm,bottom=2.5cm]{geometry}

\theoremstyle{definition}

\title{On the computation of density and two-point correlation functions of a class of random matrix ensembles}
\date{\today}
\author{Kazi Alam\thanks{kazi.a.alam@ufl.edu}, Swapnil Yadav\thanks{yadavswap.12@ufl.edu}, and K. A. Muttalib\thanks{muttalib@phys.ufl.edu}\\Department of Physics, University of Florida, Gainesville, FL 32611-8440, USA}
\begin{document}
\maketitle

\begin{abstract}
We demonstrate a method to solve a general class of random matrix ensembles numerically. The method is suitable for solving log-gas models with biorthogonal type two-body interactions and arbitrary potentials. We reproduce standard results for a variety of well-known ensembles and show some new results for the Muttalib-Borodin ensembles and recently introduced $\gamma$-ensemble for which analytic results are not yet available.
\end{abstract}

\section{Introduction}
The most common and extensively studied random matrix models are Gaussian ensembles (GE) with three different symmetries (Orthogonal, Unitary and Symplectic), and have the following joint probability distribution of eigenvalues (jpd):
\begin{align}\label{jpd}
    P(x_1,\cdots ,x_N)=C_N\prod_{j<i}^N |x_i-x_j|^\beta\prod_{i=1}^Ne^{-V(x_i)}
\end{align}
with the symmetry parameter $\beta=1,2,4$ and $V(x)=x^2$, respectively. Ensembles with $V(x)=x$, $x\ge 0$, are called Laguerre ensembles (LE). In terms of a ‘Hamiltonian’ H of the eigenvalues defined by $P=e^{-\beta H}$,  the term $\ln |x_i-x_j|$ in $H$ corresponds to a ‘two-body interaction’ of a log-gas system, while the term $\beta^{-1}V(x)$ corresponds to a single particle ‘confining potential’ (see e.g. \cite{forrester10}).
All the Gaussian (GOE, GUE, GSE) and Laguerre (LOE, LUE, LSE) ensembles are solvable through the method of orthogonal polynomials. However, for physically interesting problems, we often need to consider either a more general form of the two-body interaction, or some  special form of the confining potential, or both. Then it requires new techniques and treatments to deal with them.\\

Our interest in random matrices stems from its relation to the transport problem of disordered conductors. Parameters related to the eigenvalues of the Transfer matrix  of a quasi one-dimensional mesoscopic disordered conductor have the joint probability distribution \cite{beenakker93} 
\begin{align}\label{been}
    P(x_1,\cdots ,x_N)&=C_N\prod_{j<i}^N |x_i-x_j||s(x_i)-s(x_j)|\prod_{i=1}^Ne^{-V(x_i)}, 
\end{align}
where $s(x) = \text{sinh}^2\sqrt{x}$ and $V(x)\propto x^2$. 
The critical ensemble on the other hand has distribution of eigenvalues
\begin{align}\label{crit}
    P(x_1,\cdots ,x_N)=C_N\prod_{j<i}^N |x_i-x_j|^2\prod_{i=1}^Ne^{-V_c(x_i)}
\end{align}

where $e^{-V_c(x)}$ is the Askey weight\cite{muttalib93},  given by 
\begin{align}\label{criticalpotential}
V_c(x;q) = \sum\limits_{n=0}^\infty \mathrm{ln}[1+2q^{n+1}\operatorname{cosh}(2\operatorname{sin^{-1}}x)+q^{2n+2}], \;\;\; q <1.
\end{align}

Distribution \eqref{been} has a different two-body interaction term than the usual $|x_i-x_j|$, whereas \eqref{crit} has a different potential than standard $V(x)=x$ or $x^2$. Such variations in the jpd demand novel techniques that would allow us to solve them.\\

The purpose of the paper is three-fold. First, we demonstrate a method that allows us to compute various random matrix quantities from the joint probability distribution of a random matrix or log-gas ensemble with a general bi-orthogonal type two-body interaction and arbitrary confining potential. Second, we reproduce some known results as a verification of the method. Third, we present some new results for ensembles for which no analytical results are available.

\section{Evaluation of eigenvalue statistics}

\subsection{Biorthogonal ensembles}

We consider a generalized log-gas model\cite{muttalib95} that is a one-parameter generalization of the classical gaussian unitary random matrix ensemble (GUE) with joint probability distribution:
\begin{align}\label{bio}
    P(x_1,\cdots ,x_N)=C_N\prod_{j<i}^N |x_i-x_j||x_i^\theta-x_j^\theta|\prod_{i=1}^Ne^{-V(x_i)}
\end{align}

This is a simplified model of the joint probability distribution \eqref{been}. It was shown that the corresponding two point correlation function has a determinantal form and can be expressed as
\begin{align}\label{bioker}
    K(x,y)=e^{-\frac{V(x)+V(y)}{2}}\sum_{i=0}^{N-1}p_i(x)q_i(y),
\end{align}
where $p_i(x)$ and $q_i(x)$ are biorthogonal polynomials.
\begin{align}\label{bioortho}
    \int p_i(x)q_j(x)e^{-V(x)}=\delta_{ij}.
\end{align}

Biorthogonal ensembles reduce to classical orthogonal ensembles for $\theta=1$. Later Borodin \cite{borodin98} computed the biorthogonal polynomials, explicit form of the kernel and their asymptotics. In recent literature \eqref{bio} is known as Muttalib-Borodin (MB) ensemble \cite{forrester17}.\\

Biorthogonal polynomials were first introduced by Konhauser \cite{konhauser65} that satisfies,
\begin{align}\label{konbio}
    \int R_i(x)S_j(x)w(x)dx=\delta_{ij}.
\end{align}

$R_i(x)$ and $S_i(x)$ are made up of basic polynomials $r(x)$ and $s(x)$, respectively. Here $r(x)$ and $s(x)$ are real polynomials in x. Subscript $i$ in $R_i(x)$ or $S_i(x)$ is the degree of the polynomials, i.e., the highest power of the basic polynomial $r(x)$, $s(x)$ in $R_i(x)$, $S_i(x)$ is $i$. For Muttalib-Borodin ensemble $r(x)=x$ and $s(x)=x^\theta$. Let us denote $R_i(x)$ and $S_i(x)$ with $p_i(x)$ and $q_i(x)$ respectively. Then
\eqref{konbio} can be written as
\begin{align}
    \int p_i(x)q_j(x)w(x)dx=\delta_{ij},
\end{align}

which is same as \eqref{bioortho}.\\

Borodin computed the exact functional form of the kernel for Hermite, Laguerre and Jacobi weight, i.e., for $w_H(x)=|x|^\alpha e^{-x^2}$, $w_L(x)=x^\alpha e^{-x}$, $w_J(x)=x^\alpha$, respectively. He also calculated respective biorthogonal polynomials for each family. Finally, the limit kernels were also obtained. Limit kernel for Laguerre and Jacobi ensemble turned out to be the same. Hermite kernel is just a combination of Laguerre (or Jacobi) kernel in squared variable.
\begin{align}
    \mathcal{K}_L^{(\alpha,\theta)}(x,y)  & =\sum_{k,l=0}^{\infty}\frac{(-1)^{k+l}x^k y^{\theta l} \theta}{k!\Gamma (\frac{\alpha +1+k}{\theta}) l! \Gamma (\alpha+1+\theta l)(\alpha+1+k+\theta l)}\\
    & = \theta \int_0^1 J_{\frac{\alpha+1}{\theta},\frac{1}{\theta}}(xt) \cdot J_{\alpha+1,\theta}((yt)^\theta)t^\alpha\, dt
\end{align}

where
\begin{align}
    J_{a,b}(x)=\sum_{m=0}^\infty \frac{(-x)^m}{m!\Gamma(a+bm)}
\end{align}

is Wright's generalized Bessel function \cite{wright35}. Hermite kernel is
\begin{align}
    \mathcal{K}_H^{(\alpha,\theta)}(x,y)=\mathcal{K}_L^{(\frac{\alpha-1}{2},\theta)}(x^2,y^2) + x^\theta y\cdot\mathcal{K}_L^{(\frac{\alpha+\theta}{2},\theta)}(x^2, y^2).
\end{align}

$|xy|^{\alpha/2}\mathcal{K}_H^{(\alpha,\theta)}(x,y)$ reduces to sine-kernel and $(xy)^{\alpha/2}\mathcal{K}_L^{(\alpha,\theta)}(x,y)$ reduces to Bessel-kernel for $\theta=1$.\\

The universality of biorthogonal Laguerre kernel was also established via sine kernel and Airy kernel for the bulk and the edge, respectively, in  \cite{zhang15}. For a comprehensive review (until 2015) of MB ensembles, see \cite{forrester17}. Dolive and Tierz \cite{dolivet07} related \textit{Chern-Simons matrix models} to biorthogonal ensembles. They found a biorthogonal extension of the \textit{Stieltjes-Wigert polynomials}. Biorthogonal polynomial ensembles are also related to certain kinds of \textit{multiple orthogonal polynomials} \cite{kuijlaars10}. It turns out that a special value of $\theta$, $\theta=2$, is relevant to the random matrix model of disordered bosons \cite{lueck06}. A more general form of MB ensemble was introduced in \cite{yadav19}, called the $\gamma$-ensembles.

\subsection{Ensembles without the polynomials}

Equation \eqref{bioker} shows that any biorthogonal kernel, including classical ($\theta =1$) ones, can be calculated by evaluating the associated biorthogonal(orthogonal) polynomials. In addition, if the asymptotic limits of the polynomials are known we can also compute the limit kernel. But asymptotics for arbitrary weight is not always known. Only a few special weights are usually considered, like Hermite, Laguerre, Jacobi etc. In this section we shall show a method to handle arbitrary weight. Let us recall the biorthogonal kernel.
\begin{subequations} 
\begin{align}
	K_N^b(x,y) &= \sqrt{w(x)\,w(y)}\sum_{k=0}^{N-1} p_k(x)q_k(y).\\
	\text{set}\;\; \phi_k(x)\equiv p_k(x)&\sqrt{w(x)}\; \text{and}\; \psi_j(x)\equiv q_j(x)\sqrt{w(x)}.\\
	\text{therefore,}\;\; K_N^b(x,y) &= \sum_{k=0}^{N-1} \phi_k(x)\psi_k(y).
\end{align}
\end{subequations} 

It is possible to write $p(x)$ and $q(x)$'s in terms of basic polynomials with appropriate coefficients.
\begin{align}
    K_N^b(x,y)=\sqrt{w(x)\,w(y)}\sum_{k,l=0}^{N-1}c_{kl}\,r(x)^ks(y)^l.
\end{align}

$\displaystyle c_{kl}$ are elements of an $N\times N$ matrix. Let us denote it with $C$. We also define another $N\times N$ matrix $G$ with elements $g_{ij}$.
\begin{align}
    C=[c_{kl}]_{k,l=0}^{N-1}\,;\qquad G=[g_{ij}]_{i,j=0}^{N-1}
\end{align}

where
\begin{align}\label{gram-matrix}
    g_{ij}=\int r(x)^i s(x)^j \,w(x)\, dx.
\end{align}

It can be shown that $\displaystyle C=G^{-1}$. See \cite{borodin98, forrester10}.\\

Now the kernel becomes:
\begin{align}\label{gram-kernel}
	K_N^b(x,y)=\sqrt{w(x)\,w(y)}\sum_{k,l=0}^{N-1}[g_{kl}]^{-1}\,r(x)^k s(y)^l.
\end{align}

It provides us with a powerful tool to compute kernels straight from the jpd without going through the process of finding biorthogonal polynomials. 
(Clearly it also works for orthogonal ensembles as well since that is just a special case of biorthogonal ensembles with $r(x)=x,\; s(x)=x$.) In particular, for Muttalib-Borodin ensembles with $r(x)=x,\; s(x)=x^\theta$, it allows us to compute the kernel for arbitrary $\theta$ and arbitrary confining potential. Moreover, the generalized MB ensemble with an additional parameter $\gamma$, called the $\gamma$-ensembles \cite{yadav19}, are shown to be equivalent to  the MB ensembles with an effective $\gamma$-dependent potential. Given this effective potential, it should in principle be possible to use the present method to obtain the kernel for the $\gamma$-ensembles as well. In addition, the method works for jpd of the form (\ref{been}). It also works for a more general jpd of the form 
\begin{align}\
    P(x_1,\cdots ,x_N)=C_N\prod_{j<i}^N |r(x_i)-r(x_j)||s(x_i)-s(x_j)|\prod_{i=1}^Ne^{-V(x_i)}.
\end{align}

This technique of writing the kernel in terms of the inverted gram-matrix was used in \cite{borodin98} \footnote{Barry Simon dubbed it as \textit{ABC theorem} in \cite{simon08}}. While in Borodin's work it was used for analytic calculations available for standard linear or quadratic potentials only, our idea here is to exploit it numerically, for arbitrary potentials. Equation \eqref{gram-kernel} lets us compute many such physically interesting kernels which are not tractable analytically.    
Equation \eqref{gram-matrix} together with equation \eqref{gram-kernel} will play the central role in our computations of any kernel in the following discussion.

\subsection{Statistical analysis of level sequence}
\label{levelstat}
There are many statistical tools to investigate the distribution of eigenvalues of such log-gas or random matrix models. We will consider a few of them: global density of eigenvalues, gap functions and nearest neighbor spacing distributions(NNSD). As these are well known statistics, we will use standard definitions here. Interested readers can see \cite{mehta04} for details. The density is simply $K(x,x)$. To compare different models or kernels it's customary to \emph{unfold} the spectrum. In the unfolded variable, $x$, the density is uniform and the kernel satisfies:
\begin{align}
    K(y,y)dy=(\widetilde{K}(x,x)=1)dx.
\end{align}

Assuming the variable is unfolded we write the associated eigenvalue problem for a given kernel $K(x,y)$:
\begin{align}
    \int_{-s/2}^{s/2} \widetilde{K}(x,y)\phi(y)dy=\lambda\phi(x).\label{eigen-int-eq}
\end{align}

The gap function is given by
\begin{align}
    E_b(0;s) = \prod_{k=1}^N (1 - \lambda_k(s)).\label{0-gap}
\end{align}

For higher level gap functions \cite{mehta04, karthik16},
\begin{align}
    E_b(n;s)=E_b(0;s) \sum_{0\leq j_1<j_2<...<j_n} \frac{\lambda_{j_1}}{1-\lambda_{j_1}}...\frac{\lambda_{j_n}}{1-\lambda_{j_n}}.
\end{align}

It is straight forward to compute other functions of interest from n-level functions, see \cite[chapter 6]{mehta04}.
\begin{align}
    \label{f} F(n;s)&=-\frac{d}{ds}\sum_{j=0}^n E(j;s),\\
    \label{p} p(n;s)&=-\frac{d}{ds}\sum_{j=0}^n F(j;s).
\end{align}

$p(0,s)$ is known as \emph{nearest-neighbor spacing distribution} (NNSD). We discretize \eqref{eigen-int-eq} to compute $E_b,F$ and $p$. For the finite $N$ discrete kernel $K_N(s)$, \eqref{0-gap} reduces to a determinant
\begin{align}
	E_b^N(0;s) = \operatorname{det} (1-K_N(s)).
\end{align}

\section{Results}

Since we have obtained a general formula that should be applicable for any orthogonal or biorthogonal ensemble with arbitrary weight, the natural next step is to reproduce some known results first and then compute some new ones. In this work we consider four different classes of ensembles for verification and further investigation:

\begin{align}
    \text{Unitary Wigner-Dyson}\;&\sim\;\prod_{j<i}^N |x_i-x_j|^2\prod_{i=1}^Ne^{-V_u(x_i)},\\
    \text{Unitary critical}\;&\sim\;\prod_{j<i}^N |x_i-x_j|^2\prod_{i=1}^Ne^{-V_c(x_i)},\\
    \text{MB}\;&\sim\;\prod_{j<i}^N |x_i-x_j||x_i^\theta-x_j^\theta|\prod_{i=1}^Ne^{-V_b(x_i)},\\ \label{gamma}
    \text{$\gamma$-ensembles}\;&\sim\;\prod_{j<i}^N |x_i-x_j||x_i^\theta-x_j^\theta|^\gamma\prod_{i=1}^Ne^{-V_b(x_i)}.
\end{align}

$V_u(x)$ and $V_b(x)$ are any polynomial function. $e^{-V_c(x)}$ is the Askey weight for unitary critical ensemble \cite{muttalib93}, see (\ref{criticalpotential}). See \cite{yadav19} for $\gamma$-ensembles. Of these, the statistical properties of the first two (Wigner-Dyson and critical) are well-known, and we will reproduce them to show the validity of our method. The third one has been studied in detail only for some values of $\theta$ and only for linear and quadratic potentials, and we will verify those results. In addition we will show several new results for this model. The density of the last one has been obtained very recently for various $\gamma$, but no closed form-expression is available. We will verify some of those results as well.

\subsection{Density}

We use \eqref{gram-kernel} numerically. Appropriate scaling has to be known to achieve convergent limit quantities. 

\subsubsection{Verification of known results}

\begin{figure}[H]
  \centering
    \includegraphics[width=4.5in]{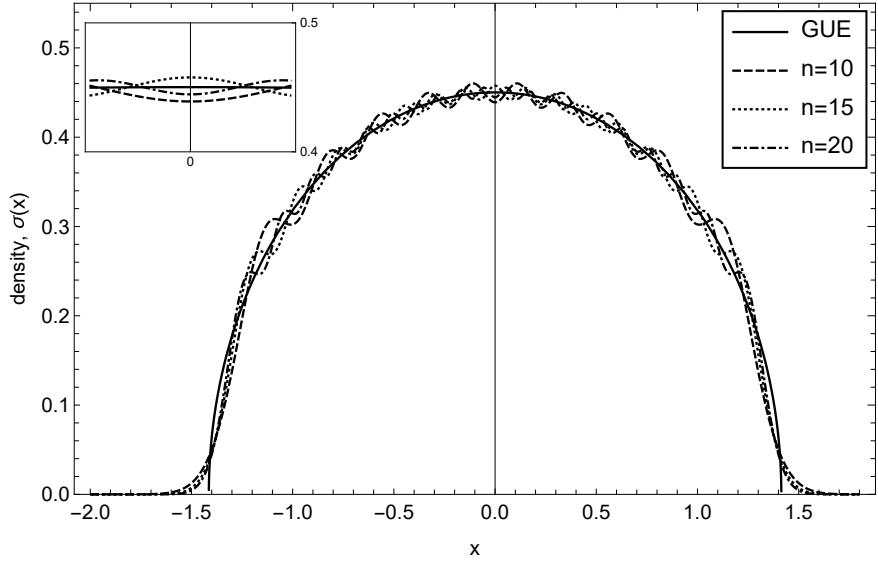}
    \caption{Convergence of the GUE density with increasing $n$. Inset shows details of the convergence near zero.}
    \label{fig:den_conv}
\end{figure}

We first use our method to obtain the density of several known results on global density that has been obtained by a variety of different methods.
To begin with, convergence of the global density to a semicircle for the GUE is shown in figure \ref{fig:den_conv}. Oscillations are due to the finite-N effect that subsides with increasing number of terms $n$.

\begin{figure}[H]
  \centering
    \includegraphics[width=4.5in]{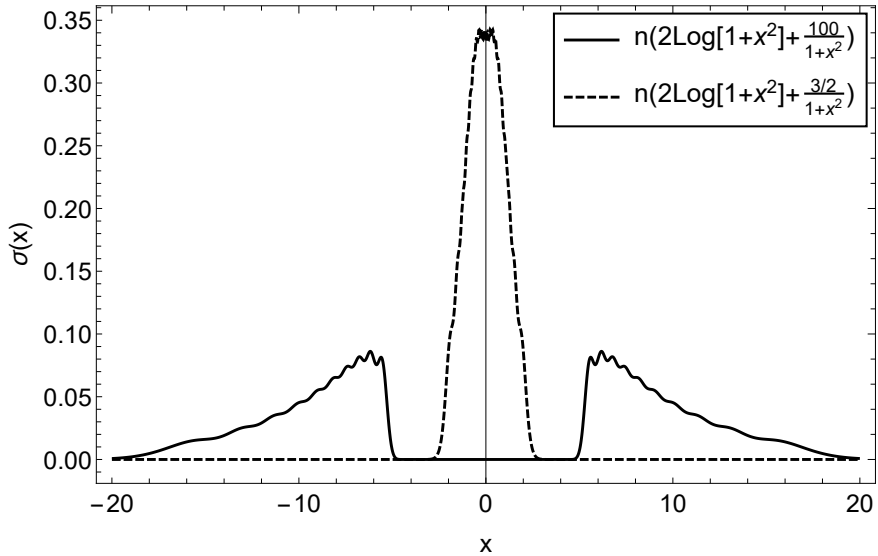}
    \caption{Density for ensemble $\prod_{j<i}^N |x_i-x_j|^2\prod_{i=1}^Ne^{-n\, \left(q\text{Log}(1+x^2)+\frac{t}{1+x^2}\right)}$.}
    \label{fig:transdensity}
\end{figure}
\begin{figure}[H]
  \begin{minipage}[t]{0.45\textwidth}
    \includegraphics[width=3in]{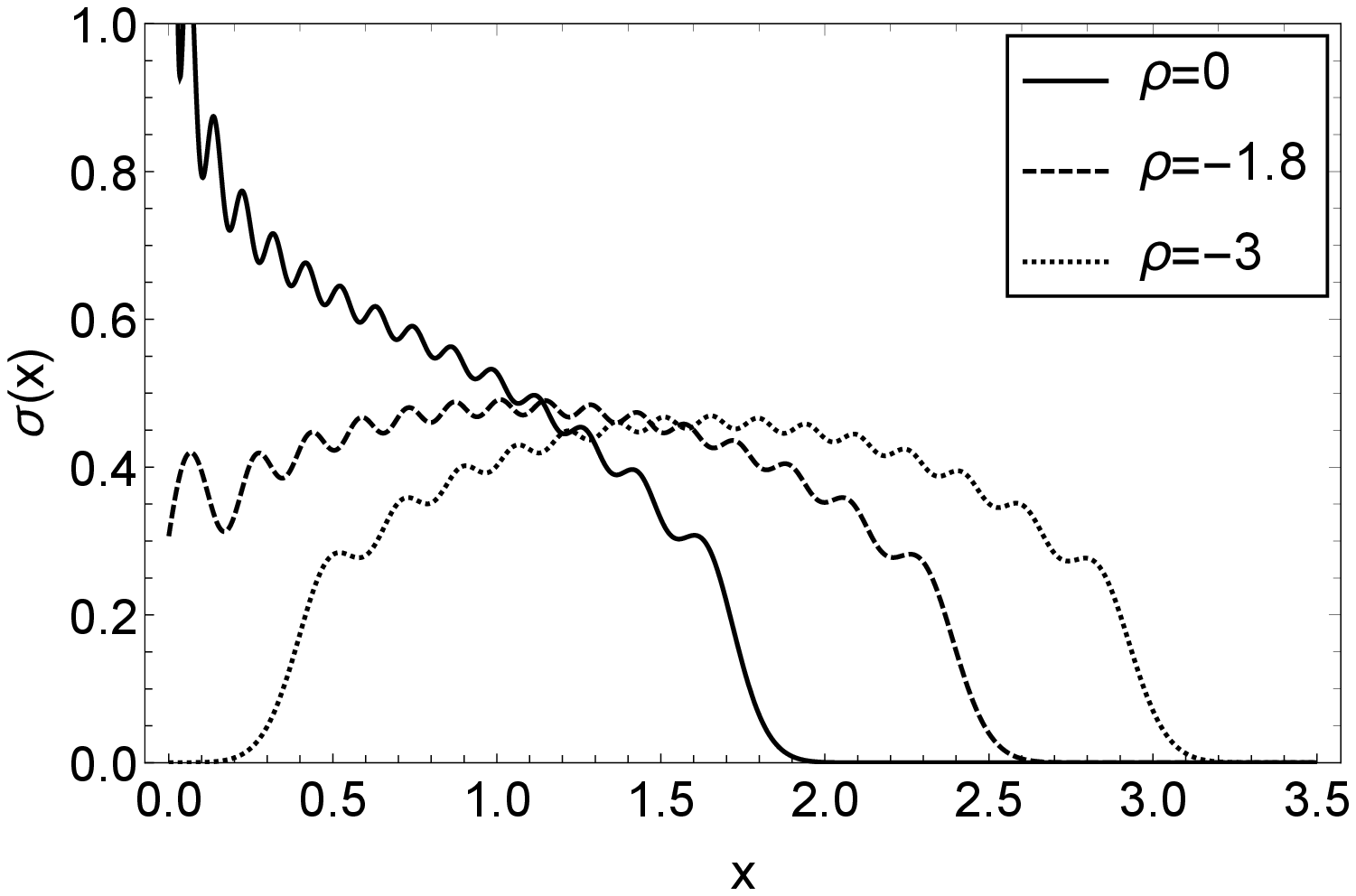}
    \caption{Transition in density of biorthogonal Laguerre ensemble for $\theta=2$ and $V(x)=x^2+\rho x$ from \protect\cite{claeys14}. First 15 terms were taken from \eqref{gram-kernel}.}
    \label{fig:crtransth2}
  \end{minipage}
  \hfill
  \begin{minipage}[t]{0.45\textwidth}
    \includegraphics[width=3in]{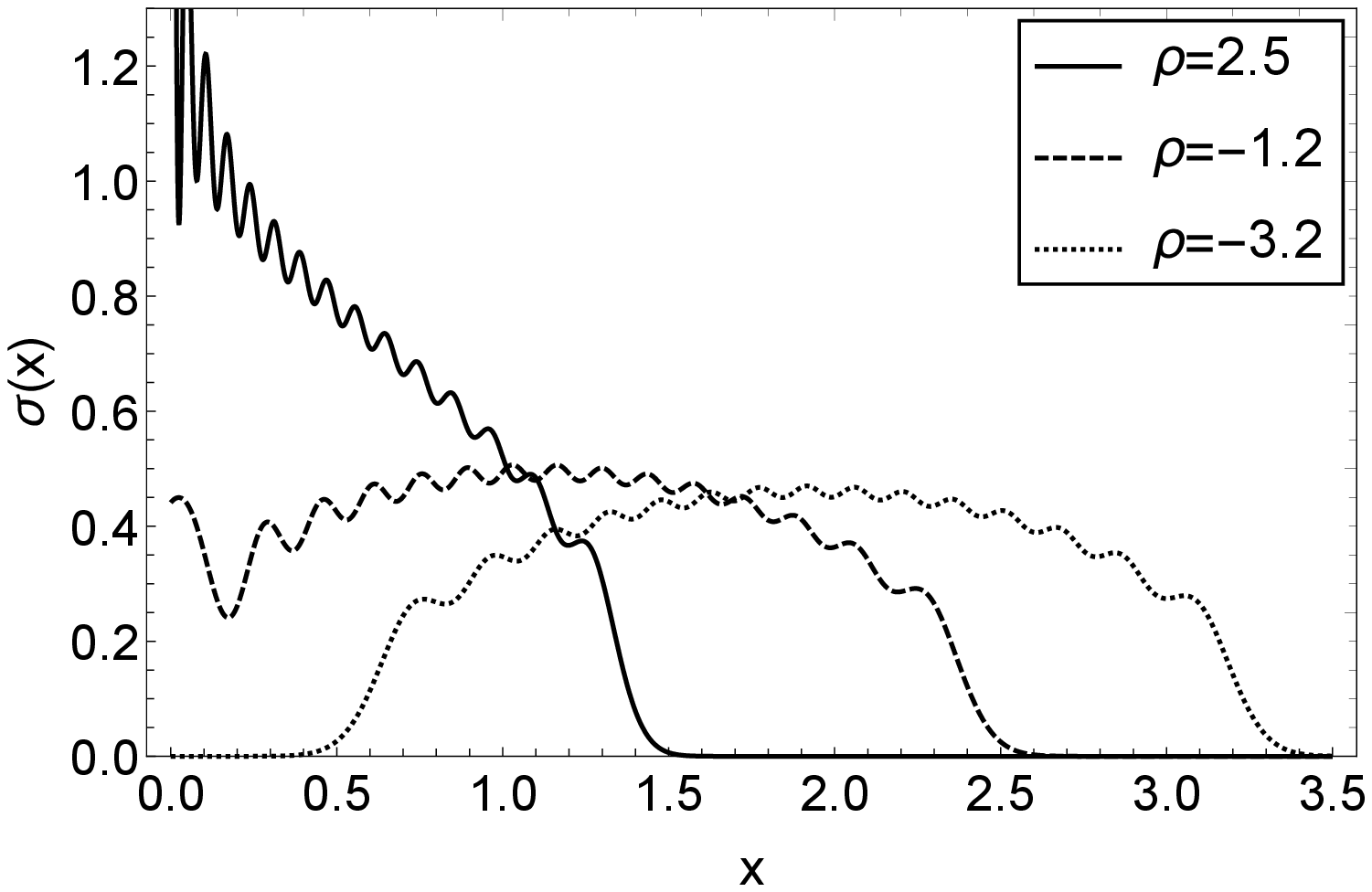}
    \caption{Transition in density of biorthogonal Laguerre ensemble for $\theta=7/2$ and $V(x)=x^2+\rho x$ from \protect\cite{claeys14}. First 15 terms were taken from \eqref{gram-kernel}.}
    \label{fig:crtransth72}
  \end{minipage}
\end{figure}

Figure \ref{fig:transdensity} shows a transition in density for an unitary ensemble with potential $\displaystyle V(x)=n\left(q\,\text{Log}(1+x^2)+\frac{t}{1+x^2}\right)$\cite{russo2020}. Another type of transition, from hard edge to soft edge, is shown in figure \ref{fig:crtransth2} and \ref{fig:crtransth72}. 

\begin{figure}[H]
  \begin{minipage}[t]{0.45\textwidth}
    \includegraphics[width=3in]{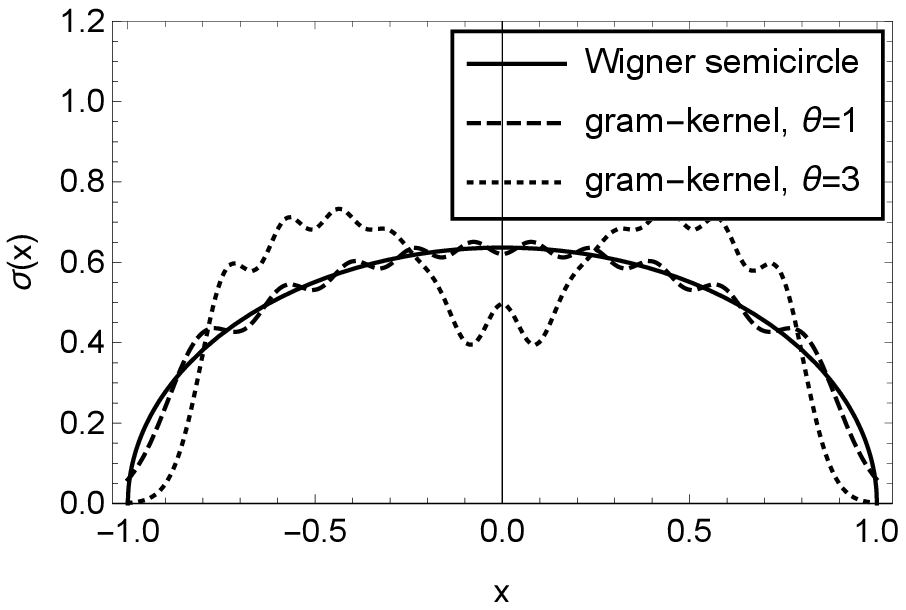}
    \caption{Density of biorthogonal Hermite ensemble for $\theta=1,3$ (wiggly lines). $\theta=1$ coincides with Wigner semicircle. First 10 terms were taken from \eqref{gram-kernel}.}
    \label{fig:biodensityh}
  \end{minipage}
  \hfill
  \begin{minipage}[t]{0.45\textwidth}
    \includegraphics[width=3in]{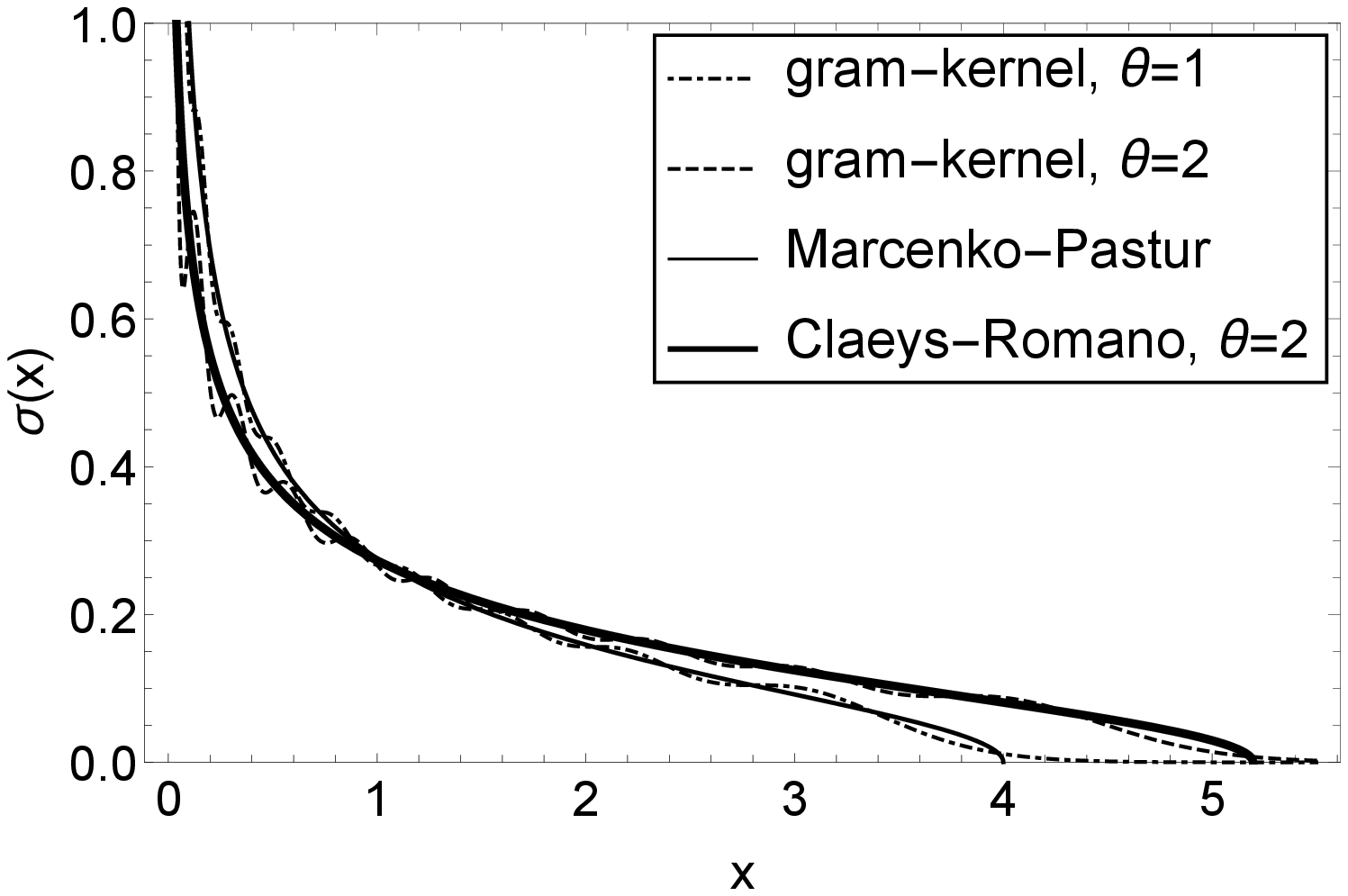}
    \caption{Density of biorthogonal Laguerre ensemble for $\theta=1,2$ (wiggly dotted lines). $\theta=1$ coincides with Mar{\v{c}}enko-Pastur (thin solid line). $\theta=2$ coincides with Claeys-Romano's analytic result (thick solid line). First 10 terms were taken from \eqref{gram-kernel}.}
    \label{fig:biodensityl}
  \end{minipage}
\end{figure}

Figure \ref{fig:biodensityh} shows global density for classical GUE($\theta=1$) and biorthogonal ensemble ($\theta=3$) along with analytic semi-circle. Figure \ref{fig:biodensityl} is the global density for similar Laguerre densities. Analytic $\theta=2$ result was taken from \cite{claeys14}.

\vspace{.5cm}

\begin{figure}[H]
  \centering
    \includegraphics[width=4in]{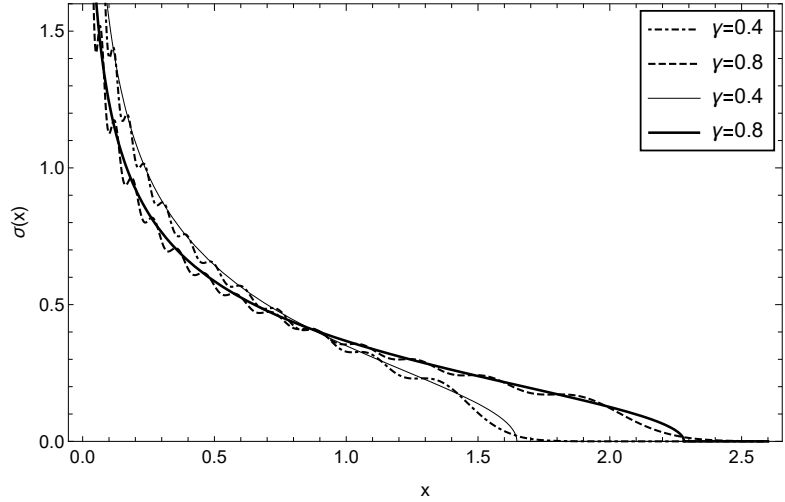}
    \caption{$\gamma$-Biorthogonal Laguerre densities for $\theta=2, \gamma=0.4,0.8$, V(x)$=2x$. Solid curves are from \protect\cite{yadav19}.}
    \label{fig:gbiodensity}
\end{figure}

Densities for $\gamma$-Biorthogonal ensemble are shown in figure \ref{fig:gbiodensity}. Effective potentials for $\gamma=0.4,0.8$ and $\theta=2$ were used from \cite{yadav19}. Potential was taken to be $V(x)=2x$. 

These results manifest the versatility of our method's applicability. We have shown that it can handle variations in the potential as well as in the repulsion term. It also works when phase transition takes place.

\subsubsection{New results on density}

\begin{figure}[H]
 \centering
    \includegraphics[width=4.5in]{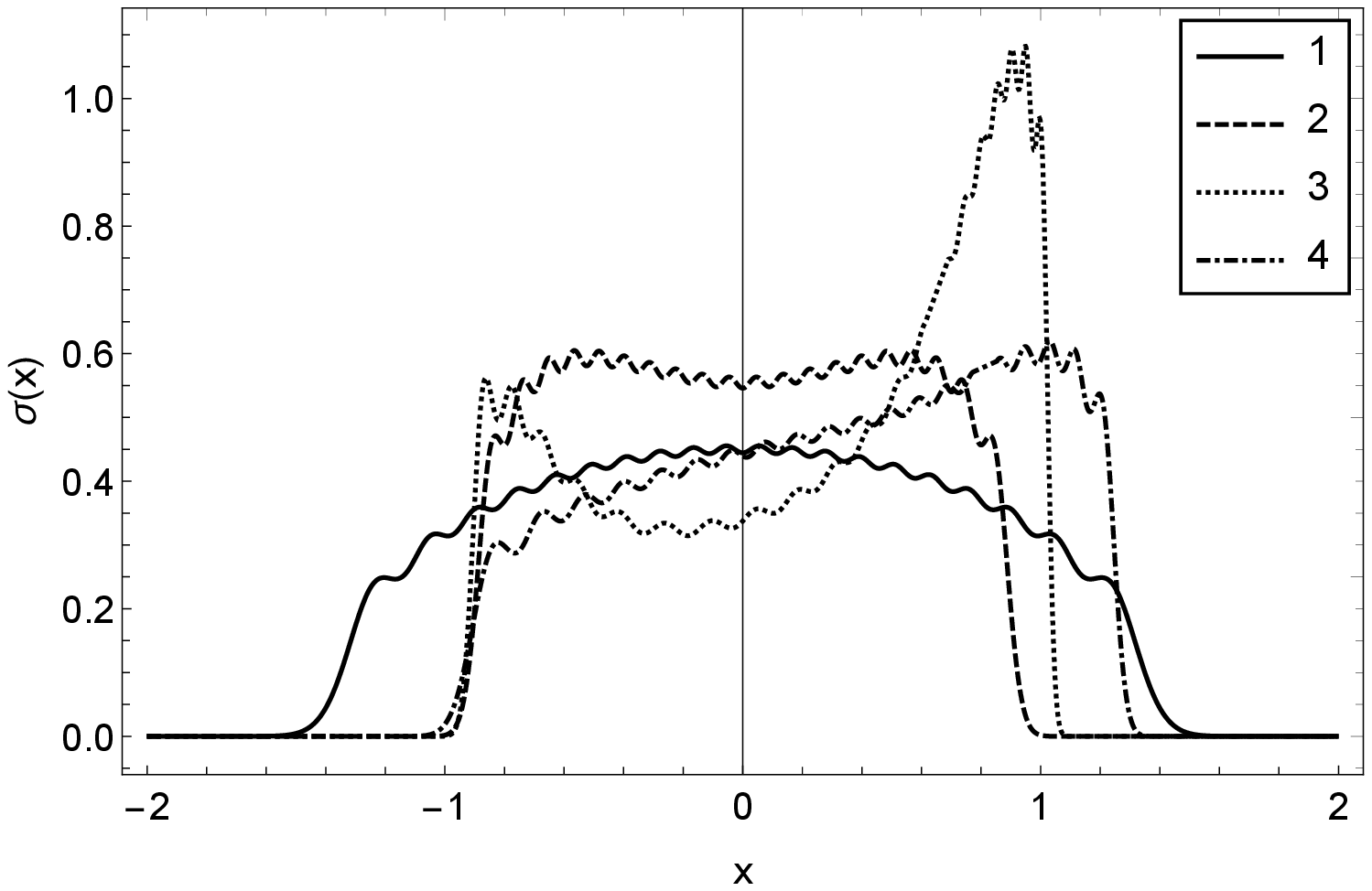}
    \caption{Densities for ensembles:\\ 1) $\prod_{j<i}^N |x_i-x_j|^2\prod_{i=1}^Ne^{-n\, x_i^2}$,\\ 2) $\prod_{j<i}^N |x_i-x_j|^2\prod_{i=1}^Ne^{-n (x_i^4+x_i^2)}$,\\ 3) $\prod_{j<i}^N |x_i-x_j||e^{2x_i}-e^{2x_j}|\prod_{i=1}^Ne^{-n\, x_i^8}$,\\ 4) $\prod_{j<i}^N |x_i-x_j||e^{2x_i}-e^{2x_j}|\prod_{i=1}^Ne^{-n\, \text{Sinh}(x_i^2)}$.}
    \label{fig:arbdensity}
  \end{figure}

In figure \ref{fig:biodensityh},  $\theta=3$ is a new result. Figure \ref{fig:arbdensity} shows densities for some ensembles with biorthogonal type repulsion and non-standard potentials. It demonstrates how the semicircle deforms into other curves due to the change in the potential and two-body interaction term, respectively. 
  \hfill

\begin{figure}[H]
  \begin{minipage}[t]{0.45\textwidth}
    \includegraphics[width=2.8in]{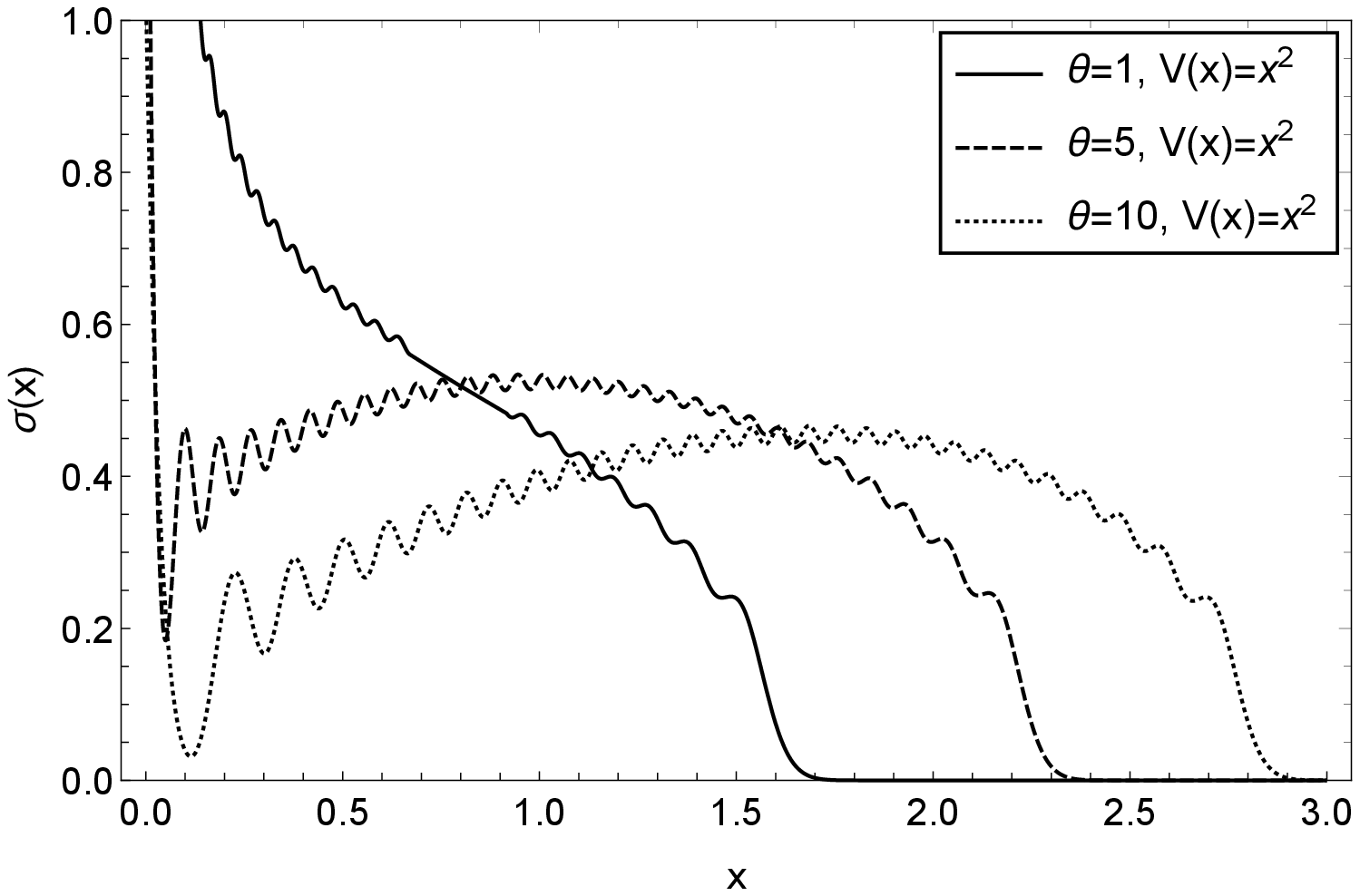}
    \caption{Density of biorthogonal Laguerre ensemble for $\theta=1,5,10$, $V(x)=x^2$. First 30 terms were taken from \eqref{gram-kernel}.}
    \label{fig:varythetadensity}
  \end{minipage}
  \hfill
  \begin{minipage}[t]{0.45\textwidth}
    \includegraphics[width=2.8in]{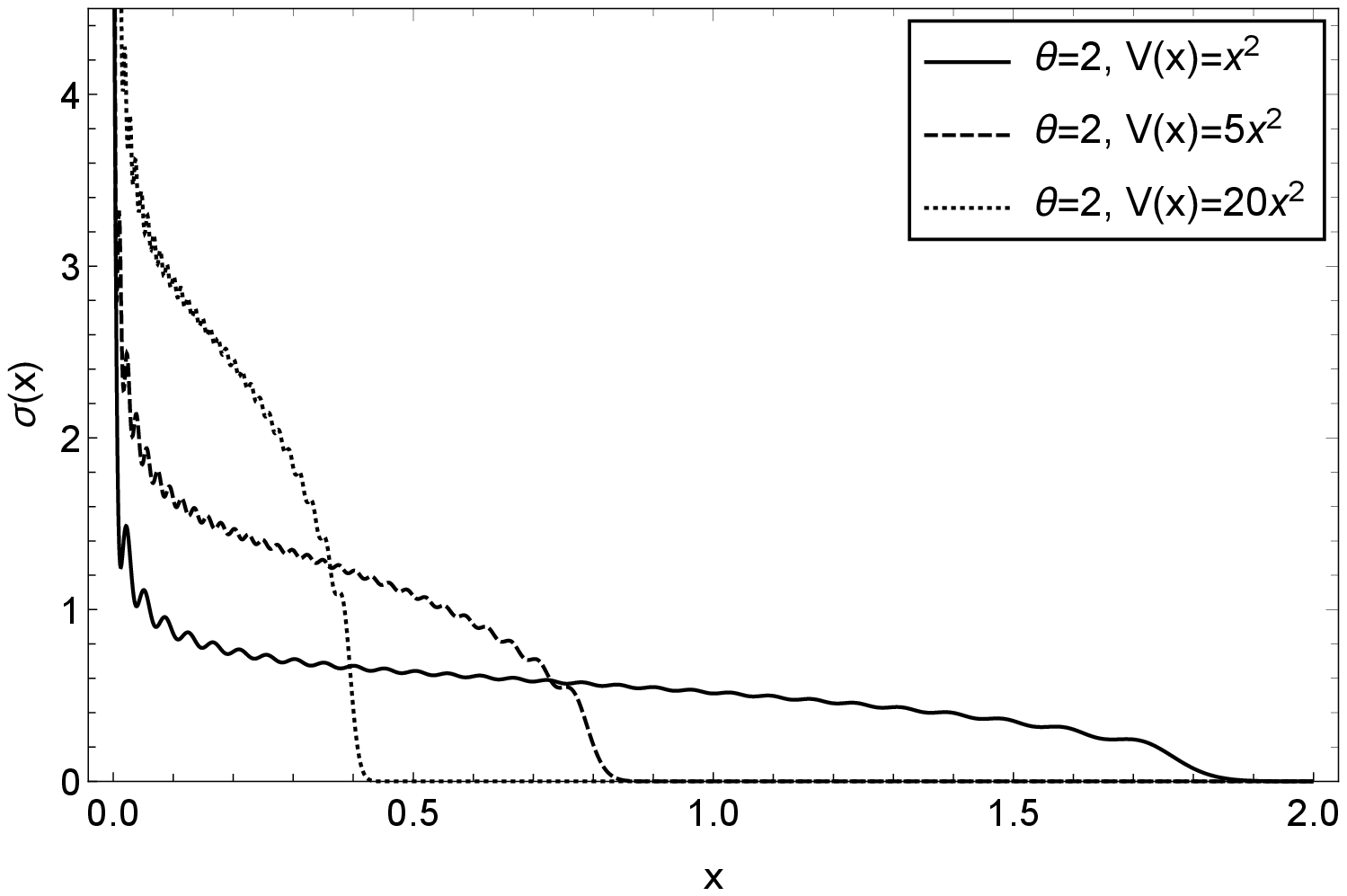}
    \caption{Density of biorthogonal Laguerre ensemble for $\theta=2$, $V(x)=x^2,5x^2,20x^2$. First 30 terms were taken from \eqref{gram-kernel}.}
    \label{fig:varypotentialdensity}
  \end{minipage}
\end{figure}

Figure \ref{fig:varythetadensity} and \ref{fig:varypotentialdensity} shows how the global density of MB ensemble changes with varying $\theta$ and the potential, respectively.

\subsection{Gap function and NNSD}

Gap function and NNSD is evaluated from the gram-kernel using formulae given in Section \ref{levelstat}.

\subsubsection{Verification of known results}

\begin{figure}[H]
  \begin{minipage}[t]{0.45\textwidth}
    \includegraphics[width=2.8in]{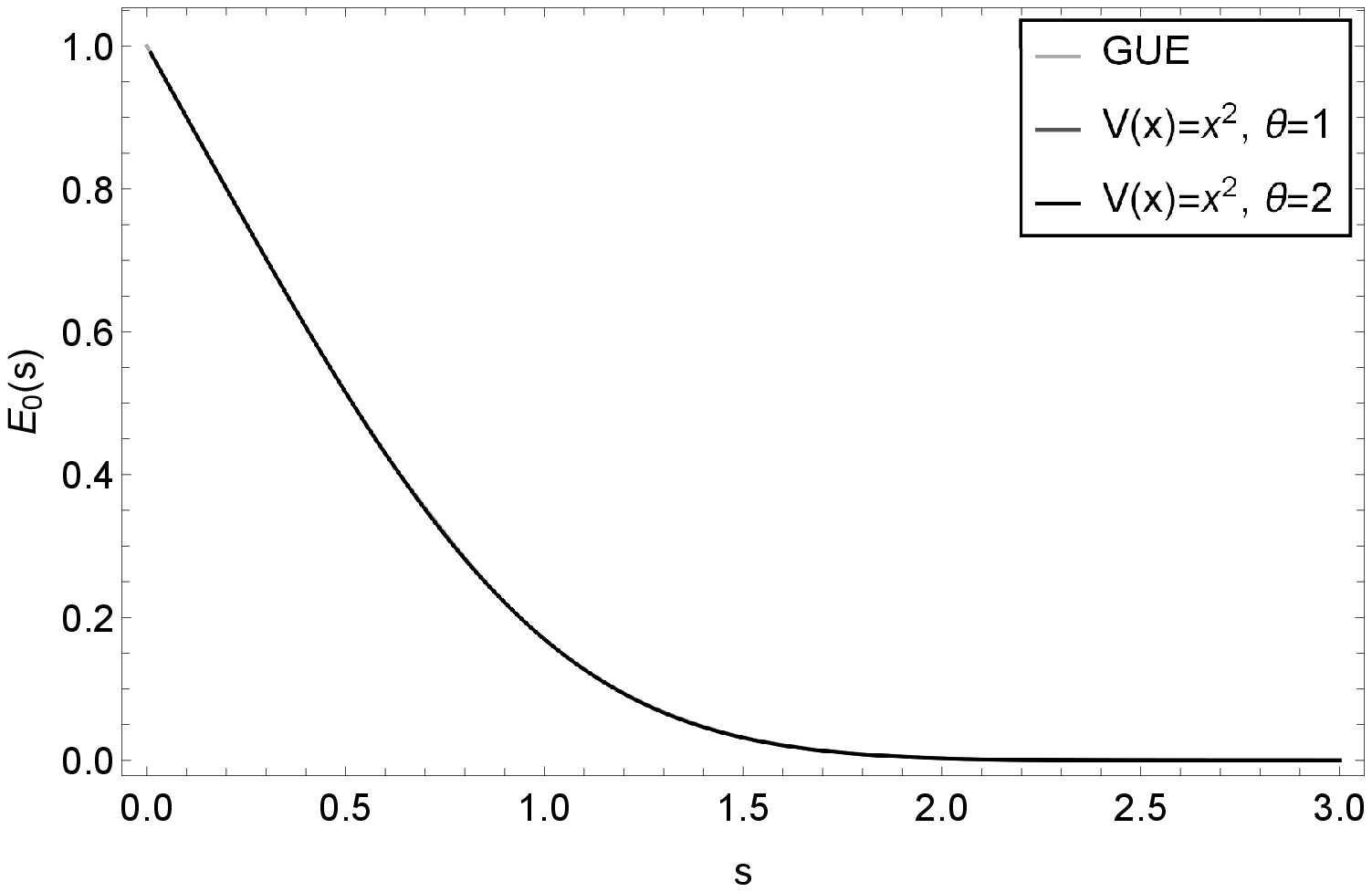}
    \caption{Gap function of biorthogonal Hermite ensemble for unfolded $\theta=1,3$, GUE and LUE with $V(x)=x^2$. Analytical data was taken from \protect\cite{mehta04}. All plots coincide. First 30 terms were taken from equation \eqref{gram-kernel}.}
    \label{fig:biogaph}
  \end{minipage}
  \hfill
  \begin{minipage}[t]{0.45\textwidth}
    \includegraphics[width=2.8in]{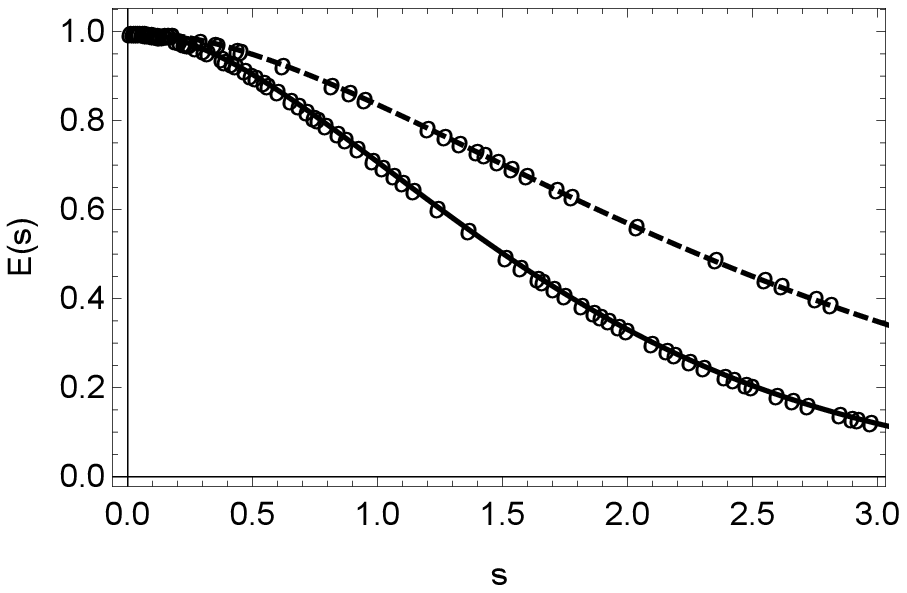}
    \caption{Gap function of biorthogonal Laguerre ensemble for $\theta=1(\text{dashed}),2(\text{solid}),\,\alpha=1$ without unfolding. Analytic results (open circles) are from \protect\cite{zhang17}. First 50 terms were taken from \eqref{gram-kernel}.}
    \label{fig:biogapl}
  \end{minipage}
\end{figure}

Gap functions in \autoref{fig:biogaph} and \ref{fig:biogapl} are drawn for biorthogonal Laguerre ensemble $\theta=1,2$ with $V(x)=x^2$. Since the analytic result was plotted in \cite{zhang17} for Laguerre, $\theta=1$ and $\alpha=1$, $\theta=2$ and $\alpha=1$, we also plotted for the same values.

Gap function in Figure \ref{fig:biogaph} was plotted for unfolded spectrum away from the edges. Unsurprisingly it coincides with GUE. This phenomenon is known as \textit{universality} in random matrix literature. Roughly, it states that, for any confining potential $V(x)$, the unitary ensembles will always produce the same gap function. Note, however, that in order to compare with available results, Figure \ref{fig:biogapl} was drawn in non-unfolded variable.\\ 

\begin{figure}[H]
  \centering
    \includegraphics[width=4in]{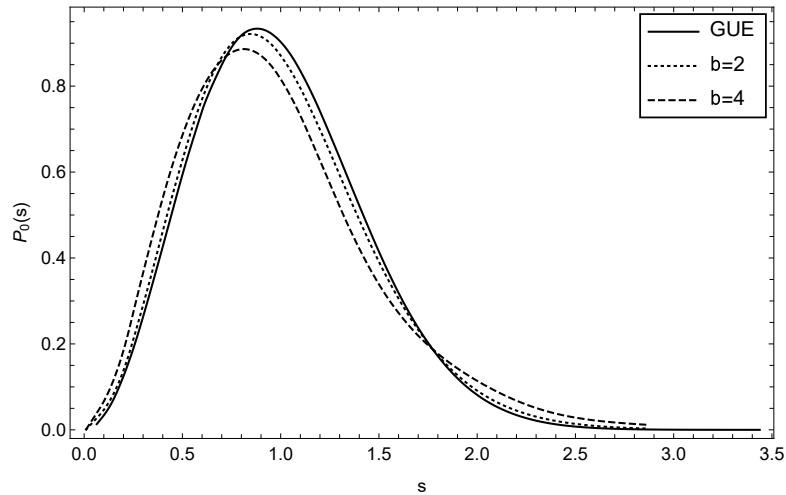}
    \caption{NNSD of critical ensemble.}
    \label{fig:critnnsd}
\end{figure}

Figure \ref{fig:critnnsd} shows the NNSD of critical ensemble. See \cite{nishigaki99} for comparison (here $b=2a$). Our result agrees well with the plots therein.

\subsubsection{New results}

\begin{figure}[H]
  \centering
    \includegraphics[width=4in]{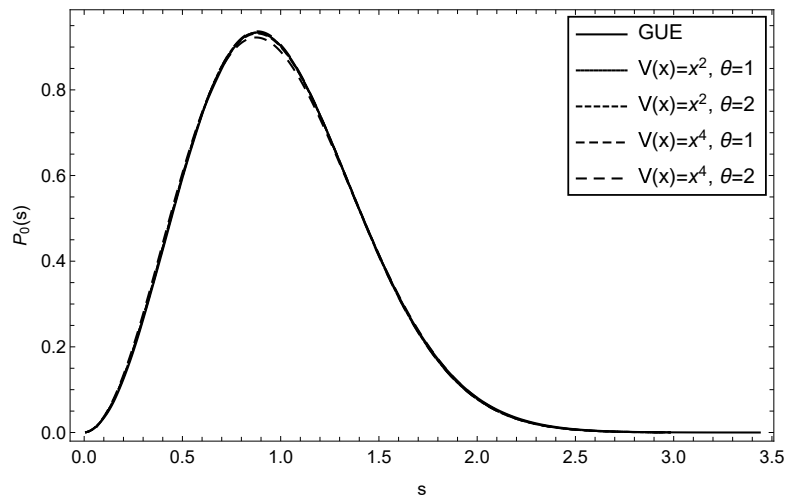}
    \caption{NNSD for GUE, biorthogonal Laguerre $\theta=1,2$ with $V(x)=x^2$ and $x^4$}
    \label{fig:bionnsd}
\end{figure}

Figure \ref{fig:bionnsd} contains NNSD for biorthogonal Laguerre ensemble $\theta=1,2$ with $V(x)=x^2,x^4$, together with Wigner-Dyson GUE. Hermite (GUE) and Laguerre distributions in the bulk overlap  due to being in the same symmetry class. $\theta =2$ also coincides. All these are new results except the GUE. Note that we called the ensembles Laguerre although the weight is different than the standard Laguerre ensemble, which is linear in $x$. Here the name Laguerre is used only to elucidate that eigenvalues range from 0 to $\infty$. The procedure we have developed can be applied for any general potential, although we showed the gap function and NNSD only for a few simple potentials.

\section{Summary and conclusion}

In this article we have shown a detailed method to obtain the kernel and other statistical quantities numerically for some general log-gas models with biorthogonal type two-body interactions and non-standard confining potentials. Universality was established for classical GUE and some general biorthogonal ensembles. 
Our goal is to study the correlations of model (\ref{been}) of a disordered quasi one-dimensional conductor within our method and then extend it to include a power $\gamma$ to the interaction, $|s(x_i)-s(x_j)|^{\gamma}$, as proposed in \cite{muttalib99,muttalib02,douglas14,muttalib05}. This is expected to give rise to a transition from metal to insulator in a disordered conductor in  three dimensions as $\gamma\le 1$ is reduced. We showed that our method is able to obtain the density for such a general gamma-ensemble as long as it can be mapped on to an MB ensemble with an effective potential. We also showed that the method remains valid when the density has a hard-edge to soft-edge transition. However, while the method allows us to obtain the two-point correlation functions of  general bi-orthogonal ensembles with arbitrary potentials, further work is needed to extend the method to obtain the two-point correlation functions of gamma-ensembles. 
This would open up the possibility of studying a new type of phase transition in disordered physical systems governed by a joint probability distribution of some relevant eigenvalues, instead of an average quantity like magnetization that acts like an order parameter.

\bibliographystyle{alpha}
\bibliography{references}

\end{document}